# Reduction of conductivity and ferromagnetism induced by Ag doping in ZnO:Co

H. Bieber[1], S. Colis[1], G. Schmerber[1], V. Pierron-Bohnes[1], D.W. Boukhvalov[2],

E. Z. Kurmaev[3], L. D. Finkelstein[3], P. Bazylewski[4], A. Moewes[4],

G. S. Chang[4], and A. Dinia[1]

[1]*Institut de Physique et Chimie des Matériaux de Strasbourg (IPCMS), UMR 7504 UDS-CNRS (UDS-ECPM), 23 rue du Loess, BP 43, F-67034, Strasbourg Cedex 2, France*

[2]*School of Computational Sciences, Korea Institute for Advanced Study (KIAS) Hoegiro 87, Dongdaemun-Gu, Seoul, 130-722, Korean Republic*

[3]*Institute of Metal Physics, Russian Academy of Sciences-Ural Division, 620041 Yekaterinburg, Russia*

[4]*Department of Physics and Engineering Physics, University of Saskatchewan, 116 Science Place, Saskatoon, SK S7N 5E2, Canada*

*Cobalt and silver co-doping has been undertaken in ZnO thin films grown by pulsed laser deposition in order to investigate the ferromagnetic properties in ZnO-based diluted magnetic materials and to understand the eventual relation between ferromagnetism and charge carriers. Hall transport measurements reveal that Ag doping up to 5% leads to a progressive compensation of the native n-type carriers. The magnetization curves show ferromagnetic contributions for all samples at both 5 K and room temperature, decreasing with increasing the Ag concentration. First principles modeling of the possible configurations of Co-Ag defects suggest the formation of nano-clusters around interstitial Co impurity as the origin of the ferromagnetism. The Ag co-doping results in a decrease of the total spin of these clusters and of the Curie temperature*

E-mail: aziz.dinia@ipcms.u-strasbg.fr (AD) and danil@kias.re.kr (DWB)

## I. Introduction

As the spin of electron entered the field of electronics, the development of semiconducting materials showing high temperature ferromagnetic properties has become an important challenge in materials science. One of the promising candidates has been shown to be ZnO doped with a transition metal such as Co or Mn. It was predicted that Mn-doped ZnO (ZnO:Mn) with a *p*-type conductivity exhibits ferromagnetic properties and a Curie temperature ($T_C$) above room temperature (RT) [1]. By extension, it was further assumed that Co doped ZnO could exhibit similar interesting properties. However, the opinions on the ZnO:Co as a diluted magnetic semiconductor (DMS) are still divided in the scientific community between the groups that are convinced of the intrinsic nature of ferromagnetism [2-4] and those who attribute the ferromagnetic properties to the presence of parasitic phases and/or structural defects [5-13]. It is, on the other hand, reported that no ferromagnetic behavior was observed in numerous thin films or bulk samples regardless of the different synthesis routes (molecular beam epitaxy [9], hydrothermal technique [14,15], co-precipitation [16-19], sol-gel [20,21], or mechanosynthesis [22]). According to X-ray magnetic circular dichroism (XMCD) study of ZnO:Co thin films exhibiting macroscopic ferromagnetic properties, no ferromagnetic signal was observed on the Co $L_3$ absorption threshold in ZnO:Co thin films in the absence of metallic Co [23,24]. Barla *et al.* noticed the absence of ferromagnetic contribution from the substitutional $Co^{2+}$ ions by XMCD and showed that the Co sublattice is purely paramagnetic [25]. These controversial results raise an issue as to whether the carrier mediated mechanism for *p*-type semiconductors may not work well for *n*-type ZnO matrix and the difficulty to achieve a *p*-type conductivity in ZnO [26,27] adds to this experimental inconsistency of the magnetic behaviors of ZnO-based DMS systems.

Many attempts have been made to obtain a *p*-type conductivity in ZnO:Co, by co-doping the system with As [28], Al [29,30], N [31], or several elements from the first group (1A) [32-35]. However, in the case of the 1A group elements with small ionic radii, the atoms have the tendency to occupy interstitial sites, thus leading to a *n*-type conductivity. In contrast, Ag, as an element of the 1B group with an ionic radius of 100 pm (larger than that of Zn), can act as an acceptor when substituting Zn ions in the ZnO matrix. It has also been suggested that Ag can have an amphoteric behavior, occupying interstitial and substitutional sites in ZnO at the same time [36]. Although this advantage makes Ag as ideal material to control the carrier density [37] of ZnO:Co, no study has focused on doping of Co:ZnO with Ag yet.

The main aim of this study is thus to explore the magnetic properties and electronic structure of Co and Ag co-doped ZnO thin films and to question the role and influence of Ag doping. This study has been carried out by employing mainly superconducting quantum interference device (SQUID) magnetometry, resonant inelastic X-ray scattering (RIXS) spectroscopy and modeling different configurations of structural defects within the density functional theory (DFT).

**II. Experimental Section**

Co and Ag co-doped ZnO targets were prepared by co-precipitation from Zn and Co acetates, and Ag nitrate dissolved in distilled water, using oxalic acid. These precursors were chosen for their high solubility in water. The 10% Co-containing $Zn_{0.9-x}Co_{0.1}Ag_xO$ powder samples with three different Ag concentrations ($x$ = 0.005, 0.01, and 0.05) were heated directly at 800 °C for 15 min. The powders were then further sedimented, pressed into the target form, and sintered at 800 °C for 30 min. The $Zn_{0.9-x}Co_{0.1}Ag_xO$ thin films were grown on $Al_2O_3$ (0001) substrates by pulsed laser deposition using a KrF excimer laser (248 nm) from Lambda Physik (LPX 100). The repetition rate and fluence were kept constant at 10 Hz and 1 $J/cm^2$, respectively. The nominal thickness was about 100 nm. The deposition temperature was set at 500 °C after preheating the substrate at 750 °C for one hour. The working gas was artificial air at a pressure of $2\times10^{-4}$ Pa. The dopant concentration in the films was estimated by electron dispersive X-ray spectroscopy (EDS) and found in fairly good agreement with the nominal values.

The magnetic hysteresis loops and the zero-field-cooled/field-cooled (ZFC/FC) magnetization curves were obtained using a Quantum Design MPMS XL SQUID magnetometer with the magnetic field applied in the film plane. All magnetization loops were corrected for the diamagnetic contribution of the substrate. The RIXS experiments which include the measurements of X-ray absorption and emission spectra (XAS and XES, respectively), were carried out on Beamline 8.0.1 of the Advanced Light Source equipped with a fluorescence endstation [38]. The emitted radiation was measured using a Rowland circle-type spectrometer. Resonant Co $L_{2,3}$ (3*d*4*s* → 2*p* transition) XES spectra were obtained at an excitation energy ($E_{\mathrm{exc}}$) near the $L_2$ absorption thresholds. On the other hand, non-resonant Zn $L_{2,3}$ and O $K\alpha$ (2*p* → 1*s* transition) XES spectra were measured at the $E_{\mathrm{exc}}$ well above the absorption thresholds. All spectra were recorded at RT and normalized to the number of photons falling on the sample monitored by the photoelectron current at a highly transparent gold mesh. The Co 2*p*, Zn 2*p*, and O 1*s* XAS spectra were measured in the total electron yield mode.

## III. Results and Discussion

The magnetization of $Zn_{0.9-x}Co_{0.1}Ag_xO$ samples ($x$ = 0.005, 0.01, and 0.05) was investigated by measuring the magnetic hysteresis loops at 5 and 300 K. The magnetic hysteresis results are presented in Fig. 1 and compared to the loop of a $Zn_{0.9}Co_{0.1}O$ reference sample without Ag doping. One can see that all samples exhibit a weak ferromagnetic contribution with the opening of the magnetization curves and a 10% remanent magnetization for $Zn_{0.85}Co_{0.1}Ag_{0.05}O$ sample at 5 K [see inset in Fig. 1(a)]. The overall magnetic properties are summarized in Table I. It should be noted that the values of both saturated magnetization ($M_S$) and remanent magnetization ($M_R$) decrease with increasing the Ag concentration. The similar trend of $M_S$ and $M_R$ variation with Ag concentration is also observed in the magnetic hysteresis loops recorded at 300 K, i.e. the sample with the lowest Ag concentration ($x$ = 0.005) has the highest $M_S$ (0.72 $\mu_B$/Co atom) and the weak ferromagnetic behavior is retained.

Three possible scenarios can be invoked to explain the observed trend of magnetic behavior of Co and Ag co-doped ZnO samples. (i) It is possible that the *n*-type carriers intrinsically present in ZnO:Co are compensated by the *p*-type doping (due to Ag) giving rise to a decrease in carrier density. (ii) Magnetic Co nano-clusters can be responsible for ferromagnetism in the films and the increase of the Ag concentration induces a decrease in the Co nano-cluster size and concentration. (iii) The presence of defects can also explain the room temperature ferromagnetism as already proposed by several groups [39-41]. In the same vein, it is possible that the increase in Ag concentration is accompanied by a decrease in the defect concentration.

To assess the above suppositions, additional investigations using Hall transport and ZFC/FC magnetization measurements under different applied magnetic fields were carried out. The results of Hall transport measurements at 77 and 300 K are presented in Table II. Both the undoped and Ag-doped ZnO:Co samples exhibit *n*-type conductivity. The carrier concentration was found to decrease from $4.2\times10^{15}$ down to $7.9\times10^{10}$ $cm^{-3}$ at 77 K and from $1.2\times10^{18}$ to $3.0\times10^{14}$ $cm^{-3}$ at 300 K with an increase in Ag doping of up to 5 %. This clearly indicates that positive charge carriers resulting from Ag doping compensate the native *n*-type conductivity of ZnO although the Ag doping up to 5 % is not sufficient to lead to *p*-type conductivity. The carrier densities of all samples at both 5 and 300 K are in the insulating regime (less than $2\times10^{18}$ $cm^{-3}$), where the variable range hopping process governs the conduction at low temperature. In this regime, the magnetic properties of ZnO:Co samples cannot be associated with the carrier mediated mechanism which requires a carrier concentration above $10^{20}$ $cm^{-3}$ for the onset of ferromagnetism [1].

Secondly, the presence of magnetic clusters in Co and Ag co-doped samples was investigated via the ZFC/FC magnetic measurements. Fig. 2 shows the ZFC/FC curves of $Zn_{0.9-x}Co_{0.1}Ag_xO$ samples with $x$ varying from 0 to 0.05. For the ZFC measurements, the samples were cooled down to 5 K under zero magnetic field and then the magnetization was measured while heating up to 300 K in the magnetic field ($H_{app}$) of 5 $10^{-2}$ or $10^{-1}$ T applied along the film plane. For the FC measurements, the $H_{app}$ was fixed at a given value during the cooling procedure, followed by the magnetization measurements at the same field value while heating up the sample to 300 K. The variation of ZFC/FC magnetization depends on both Ag concentration and applied field. The overall results show, for all samples, the existence of a critical temperature below which the ZFC and FC curves do not superimpose. This is directly correlated with the presence of magnetic Co nanoparticles diluted in the ZnO matrix. The ZFC/FC magnetization variations can be explained by the coexistence of weak ferromagnetic and superparamagnetic phases with a transition temperature depending on the particle size. The temperature-dependent magnetic behavior of metallic Co particles with a size of few nm is characterized by the change from ferromagnetic to super-paramagnetic states around the blocking temperature ($T_B$). At high temperature, the magnetic moment of each nanoparticle rapidly fluctuates whereas below $T_B$ the magnetic moments appear as "blocked". The $T_B$ depends on the measurement typical time ($t_m$) [42]:

$$T_B = \ln\left(\frac{t_m}{\tau_0}\right)^{-1} \times \frac{K_a V}{k_B} \qquad (1),$$

where $V$ is the volume of particles, is a length of time, characteristic of the material, $\tau_0$ - called the attempt time or attempt period (its reciprocal is called the attempt frequency); its typical value is $10^{-9}$–$10^{-10}$ second, $k_B$ is the Boltzmann constant, and $K_a$ is the effective anisotropy constant (i.e. magneto-crystalline constant) which is equal to 0.45 MJ/m$^3$ for hcp Co. Following the work of Chantrell *et al*. [42] and Denardin *et al*. [43], to have information on the nanoparticles size distribution in our thin films, we deduced the blocking temperature distribution from the difference between the FC and ZFC magnetization curves for all samples as shown in Fig. 2 according to the following equation [see inset of Fig. 2(c)]:

$$\mu_{(tot)}^{(ZFC)} - \mu_{(tot)}^{(FC)} = k \int_T^{\infty} T_B \rho(T_B)\, dT_B \qquad (2),$$

where $\kappa$ is a constant. After fitting this distribution to a lognormal distribution:

$$\rho(T_B) = \frac{1}{(\sigma\sqrt{(2\delta)})}\frac{1}{T_B}\exp\left(\frac{(-\ln(T_B/T_{(B,C)})^2)}{(2\delta^2)}\right) \qquad (3),$$

(where $\sigma$ is the standard deviation in logarithm) the following median temperatures of $Zn_{0.9-x}Co_{0.1}Ag_xO$ can be obtained: $T_{B,C}$ = 3.6 K ($x$ = 0), 2.1 K ($x$ = 0.005), 5.9 K ($x$ = 0.01), and 5.2 K ($x$ = 0.05). According to Eq. (1), these median temperatures correspond further to the mean diameter of Co nanoparticles of $D$ = 1.80(5) nm, 1.47(5) nm, 2.10(5) nm, and 2.01(5) nm for the undoped and Ag-doped ZnCoO thin films with $x$ = 0.005, 0.01 and 0.05, respectively. These results are in agreement with tomographic measurements in ZnO:Co system where Co clusters with few nm in size are observed[36]. It is important to notice that these authors have used an effective anisotropy constant of fcc Co, which is about ten times smaller than the value for hcp Co resulting into a larger nanoclusters size. This clearly reflects that, although Co nanoparticles are present in the samples studied here, their size distribution is irrelevant to the Ag doping as shown in the inset of Fig. 2(c). Therefore, the presence of metallic Co nanoparticles can be ruled out as the origin of the observed room temperature ferromagnetic behavior of Co and Ag co-doped ZnO thin films.

In view of the magnetic hysteresis and ZFC/FC magnetization measurements, several magnetic contributions can be addressed. Beyond the paramagnetic contribution of $Co^{2+}$ ions substituting for Zn ions [20, 25] and considering the results of the magnetic measurements, the presence of metallic Co clusters [44] has been confirmed in all the studied samples. The size and the distribution of Co particles cannot explain the room temperature ferromagnetism in the undoped and Ag-doped ZnO:Co thin films. In addition, the presence of Ag around Co ions in the ZnO matrix can favor the bonding between the Co ions through O neighbors, leading to an antiferromagnetic superexchange mechanism. Therefore, the decrease in $M_S$ with Ag doping suggests an increase of the antiferromagnetic coupling between these Co ions as the Ag concentration increases. The role of the charge carriers is also questionable, as the most resistive sample, where the compensation of electrons by the 5% Ag doping is maximal, exhibits a ferromagnetic signal. It is therefore possible that the ferromagnetic contribution, which increases upon decreasing the Ag concentration, originates from the presence of defects inside the ZnO matrix. It seems therefore that the ferromagnetic contribution is related to structural defects such as Zn interstitials or O vacancies as previously reported [11-13,39-41]. These defects, which were claimed to induce ferromagnetic properties even in pure and usually nonmagnetic oxides such as ZnO, $TiO_2$, etc., are also present in ZnO:Co thin films giving rise to a ferromagnetic signal of the same order of magnitude as in the samples with additional Ag-doping. However, we observe a decrease of the ferromagnetic contribution related to defects at room temperature with Ag-doping. This can be explained by an enhancement of the crystalline quality of ZnO:Co thin films with Ag-doping as suggested by X-ray diffraction measurements (not reported here). The full width at half maximum of the rocking curve recorded on the ZnO(0002) peak decreases with the Ag-doping (from 1.3° in the undoped ZnCoO film down to 0.66° for the 5% Ag-doped thin film). Moreover, it is well known that well-crystallized single-phase bulk materials and uniform defect-free thin films exhibit no (or negligible) ferromagnetic signal [45,46]. Theoretical models, describing the role of defects on the magnetic properties in DMS systems, show that small defect concentration favors the formation of an impurity band in the ZnO band gap. The spin-splitting of this impurity band leading to a ferromagnetic signal can occur in a spontaneous way or can be due to the presence of further impurities or magnetic defects [47]. However, the control of ferromagnetic properties due to defects is still very difficult to achieve experimentally.

In order to have a better insight on the electronic properties of undoped and Ag-doped ZnO:Co films, X-ray spectroscopy measurements were carried out. One of the main advantages of X-ray spectroscopy for the study of DMS systems is the possibility to selectively excite and measure the electronic structure of impurity atoms. With the advent of synchrotron sources of the 3rd generation, it is possible to obtain intense not only X-ray absorption spectra but also fluorescent X-ray emission spectra, and thus to access unique information about local bonding of impurity atoms. Fig. 3 shows the O $K\alpha$ and Zn $L_3$ XES spectra of Co and Ag co-doped ZnO films which probe the occupied O 2*p* and Zn 3*d* densities of states, respectively. The Fermi level ($E_F$) on the spectral curves is determined from X-ray photoelectron spectroscopy measurements which allow to obtain the O 1*s* and Zn 2*p* binding energies of ZnO:Co samples [48]. It is obvious that the Zn 3*d* states reside at the bottom of the valence band (feature *A*) where they hybridize with the O 2*p* states. On the other hand, O 2*p* states are concentrated at the top of the valence band (feature *B*) where they hybridize with the Zn 3*d* and 4*p*-states. For both spectra, there is no noticeable change in the vicinity of $E_F$. This reflects that Ag doping does not induce any redistribution of O 2*p* and Zn 3*d* states in the valence band.

Zn 2*p* XAS spectra (2*p* → 4*s*4*d* transition), which probe unoccupied Zn 4*s* and 4*d* states, show a systematic increase of the intensity for the near-edge peak (located at about 1024 eV) with Ag-doping [see Fig. 4 (a)]. The spectral change is more prominent for 5% Ag doped ZnO:Co sample which contains 10 times more dopants than $Zn_{0.895}Co_{0.1}Ag_{0.005}O$ sample. Ag 2*p* XAS (2*p* → 5*s*5*d* transition) of $Ag_2O$[41] and Co 2*p* XAS (2*p* → 4*s*3*d* transition) of $Zn_{0.895}Co_{0.1}Ag_{0.005}$ are shown in Fig. 4 (b-c); all spectra are aligned by the onset of absorption. This comparison shows that an increase of the near-edge peak of Zn 2*p* XAS is caused by hybridization of Zn 4*d*, Co 3*d* and Ag 5*d*-states.

Fig. 5 presents the Co $L_{2,3}$ XES spectra resonantly excited at $L_2$-threshold ($E_{exc}$ = 794.7 eV). This threshold excitation creates the favorable conditions for the excitation of Co $L_2$-level with respect to that of Co $L_3$-level. As seen, the I($L_2$)/I($L_3$) intensity ratio decreases with increasing Ag content which can be attributed to the suppression of magnetic contribution to this ratio according to theoretical calculation [49]. Ag doping reduces the magnetic moment on Co atom as the I($L_2$)/I($L_3$) intensity ratio depends on the spin polarization of occupied and unoccupied *d*-states as formulated in the following [50]:

$$\frac{I(L_2)}{I(L_3)} = \frac{\Gamma_3}{2\Gamma_2} \frac{N\bar{N} + (1/9)M\bar{M}}{N\bar{N} + (5/9)M\bar{M}} ,$$

where $\Gamma_2$ ($\Gamma_3$) is the full width at half maximum of $L_2$ ($L_3$) level, $N$($\bar{N}$) is the number of occupied (unoccupied) *d*-states in exciting atom (Co in our case) and $M$ ($\bar{M}$) is the spin polarization of occupied (unoccupied) *d*-states. Using this formula, one can obtain the reduction of I($L_2$)/I($L_3$) intensity ratio along with a decrease of magnetic moment for 3*d*-monoxides [50].

The modeling was performed by density functional theory (DFT) using the SIESTA pseudopotential code [51], as done in our previous works [20]. All calculations were performed using the generalized gradient approximation (GGA-PBE) [52]. A full optimization of the atomic positions was performed. During optimization, the electronic ground state was consistently found using norm-conserving pseudo-potentials for cores, a double-ζ plus polarization basis of localized orbitals for zinc, silver, and oxygen. Cut-off radii of the pseudopotentials is 2.50, 2.40. 2.30 and 1.50 for Ag, Zn, Co and O respectively. Optimization of the force and total energy was performed with an accuracy of 0.04 eV/Å and 1 meV, respectively. All calculations were carried out with an energy mesh cut-off of 360 Ry and a *k*-point mesh of 4×4×4 in the Mokhorst-Park scheme [53]. The formation energies ($E_{\mathrm{Form}}$) were calculated using the standard method described in details in Ref [20]. The Curie temperature ($T_C$) was estimated using the formula $T_C = JS(S+1)/3k_B$ proposed for the Ising model, where $J$ is the exchange interactions defined for the Hamiltonian $H_{ij} = -J_{ij}S_iS_j$, $S$ the total spin of defect-clusters, and $k_B$ the Boltzmann constant. To understand the nature of ferromagnetism in the ZnO:Co, different combinations of structural defects were calculated in a ZnO matrix. Taking into account our previous computer modeling of Co impurities in semiconductors [20], we have performed calculations for; (i) a single substitutional Co impurity for a Zn atom ($Co_S$), (ii) a pair of substitutional Co impurities ($2Co_S$), and (iii) various defect-clusters consisting of Co atoms at the cation sites and interstitials ($x$$Co_S$+$Co_I$) as depicted in Fig. 6. The realistic impurity concentrations were simulated by creating the supercells consisting of 128 atoms of Zn and O for clusters containing up to 5 Co atoms [Fig. 7(a)], and 256 atoms of Zn and O for clusters containing more than 6 Co atoms [Fig. 7(b)]. The concentration of Co impurity atoms in both cases was about 5~7 %. By doubling the supercell along the *y*-axis, the difference of total energy between ferromagnetic and antiferromagnetic orientations of the total spin on defect-clusters was calculated in order to simulate the possible long-range magnetic ordering. On the other hand, the exchange interactions within the cluster were obtained by calculating the difference in total energy between ferromagnetic and antiferromagnetic spin orientation of Co atoms in the cluster.

The simulation results are summarized in Table III. As is evident from these results, the substitutional Co atoms for the cation sites are coupled only by antiferromagnetic exchange interaction. The presence of O vacancies ($V_O$) near the substitutional Co atoms (e.g. $2Co_S+V_O$) under these realistic impurity concentrations does not induce a long-range magnetic order. However, the configuration of defect-clusters involving both interstitial and substitutional Co atoms [Figs. 6(e) and (f)] gives rise to increasing charge carriers in the system and the appearance of RKKY-type long-range exchange interaction [54]. In this case, an increase of the number of $Co_S$ leads to a decrease in the formation energy and an increase of exchange interactions between the clusters ($J_2$) as well as within clusters ($J_1$) as presented in Table III. The clusters of $3Co_S+Co_I$ are found to be energetically the most favorable ($E_{Form}$ = 1.07 eV) and then the formation energy increases again with increasing the number of $Co_S$ atoms.

When all 6 Zn atoms around the interstitial Co impurity are substituted with Co atoms at the concentration about 5~7 % ($6Co_S+Co_I$), the inter-cluster distance also increases up to 2 nm, resulting in a substantial reduction of the exchange interaction (close to 0). The large $6Co_S+Co_I$ clusters with about 0.5 nm in size are energetically less favorable, but can be possibly formed in the thin films. This is consistent with our magnetic hysteresis and ZFC/FC magnetization measurements. The metallic Co nanoclusters with a mean diameter about 2 nm in $Zn_{0.9-x}Co_{0.1}Ag_xO$ samples correspond to the inter-cluster distance longer than 3 nm. That is why the long-range magnetic ordering could not be observed in the experiment. Based on the calculated formation energies and exchange interactions, we can conclude that $xCo_S+Co_I$ clusters (most likely $3Co_S$+$Co_I$) are responsible for ferromagnetism in ZnO:Co for Co concentrations of about 10%.

Taking into account that the O vacancies are considered as one of the possible origin of ferromagnetism in ZnO:Co system [41,55], we have performed the calculation for the combinations of metal impurities and O vacancies. The formation energy of a single O vacancy equals to 4.12 eV, which is close to the previous calculations [55]. Therefore, just as in the case of a combination of metal impurities, the location of the oxygen vacancies is energetically the most favorable near impurities. The calculation results presented in Table IV indicate that the presence of O vacancies produces very little effect on the magnetic properties of the system, but increases the formation energy. The ferromagnetic interaction between distant defects (corresponding to a low defect concentration) occurs only in the case of the formation of nano-clusters consisting of a small number (up to 4 or 5) of Co atoms located not only in cation sites, but also in the interstitial sites. It is worth noting two important points: i) the $3Co_S$+$Co_I$ clusters form the ferromagnetic network and thus a stable magnetism with respect to temperature fluctuations, ii) ferromagnetic network exists on a paramagnetic background, created by single Co-atoms occupying cation sites (having the lowest formation energy) and the large Co-clusters dispersed in the host matrix. We have compared the Co 2*p* XAS spectra of $Zn_{0.9-x}Co_{0.1}Ag_xO$ system with those of CoO and Co reference samples as presented in Fig. 8. It is seen that the spectra of $Zn_{0.9-x}Co_{0.1}Ag_xO$ thin films show intermediate situation between that of metallic Co and of the Co oxide. The absorption line is narrow and the multiplet structure at the low-energy side disappears. This also supports the idea of Co defect-cluster formation together with the uniform distribution of Co atoms occupying Zn-sites.

Finally, it is necessary to examine the issue of the reduction of Co magnetic moment under Ag co-doping as confirmed from the magnetic hysteresis measurements and DFT calculations. When the $Ag^{1+}$ impurity substitutes for $Zn^{2+}$ sites near the $Co_I$ impurity, the substitution leads to the capture of missing electron from $Co_I$ atoms to satisfy the charge neutrality of system. This results in the reduction of total magnetic moment of Co cluster [in the selected arrangement of spins for $3Co_S$+$Co_I$ configuration as depicted in Fig. 6(e)], *i.e.* the decrease of both Co magnetic moment and the number of charge carriers. The decrease in carrier concentration is accompanied by the suppression of ferromagnetic exchange interactions inside and between the defect-clusters also causing the decrease of $T_C$ from 511 K for $3Co_S$+$Co_I$ cluster to 372 K for $3Co_S$ + $Co_I$ + $Ag_S$. Therefore, the modeling of clusters such as $3Co_S$+$Co_I$+$Ag_S$ with antiferromagnetic exchange interaction between $Co_S$ and $Co_I$ allows to explain the appearance of ferromagnetism in Co and Ag co-doped ZnO DMS system with a good consistency between experimental and calculated magnetic moments and $T_C$.

## IV. Conclusion

To conclude, we have reported the grown of the Co and Ag co-doped ZnO thin films using a pulsed laser deposition and investigated the influence of Ag doping on the magnetic properties of ZnCoO diluted magnetic semiconductors. The Hall effect measurements have shown a progressive decrease of *n*-type carriers due to the introduction of Ag impurities. Magnetization measurements have shown a ferromagnetic contribution at room temperature which decreases upon Ag doping. The overall results of the present study support that ferromagnetic contribution is mainly due to the formation of defects. This is also supported by our DFT calculations suggesting that the presence of Co impurity atoms at the interstitial sites ($Co_I$-type defect) plays a key role in the formation of long-range ferromagnetic ordering by creating the electron carriers and participating to the formation of magneto-active cluster $3Co_S$+$Co_I$. The Ag impurity within the $3Co_S+Co_I+Ag_S$ cluster is responsible for the reduction in charge carriers and magnetic moment on Co ions as well as weakening of magnetic behaviors.

## Acknowledgements

We gratefully acknowledge the Natural Sciences and Engineering Research Council of Canada (NSERC) and the Canada Research Chair program. This work is also partly supported by the Russian Science Foundation for Basic Research (Projects 11-02-00022). H.B. would like to thank the Region Alsace for financial support. DWB thank KIAS Center for Advanced Computation for providing computing resources.

**Table I.** Saturation magnetization ($M_S$), remanent magnetization ($M_R$), and coercive field ($H_C$) values taken from the magnetization measurements at 5 and 300 K.

| Samples | Magnetic properties at 5 K | | | 300 K | | |
|---|---|---|---|---|---|---|
| | $M_S$ ($\mu_B$/Co atom) | $M_R$ ($\mu_B$/Co atom) | $\mu_0 H_C$ (T) | $M_S$ ($\mu_B$/Co atom) | $M_R$ ($\mu_B$/Co atom) | $\mu_0 H_C$ (T) |
| $Zn_{0.9}Co_{0.1}O$ | 1.29 | 0.17 | $2.2\ 10^{-2}$ | 0.55 | 0.03 | $0.6\ 10^{-2}$ |
| $Zn_{0.895}Co_{0.1}Ag_{0.005}O$ | 1.20 | 0.12 | $1.8\ 10^{-2}$ | 0.72 | 0.03 | $0.6\ 10^{-2}$ |
| $Zn_{0.89}Co_{0.1}Ag_{0.01}O$ | 1.00 | 0.08 | $3.4\ 10^{-2}$ | 0.24 | 0.04 | $1.4\ 10^{-2}$ |
| $Zn_{0.85}Co_{0.1}Ag_{0.05}O$ | 0.83 | 0.02 | $1.6\ 10^{-2}$ | 0.09 | 0.01 | $0.8\ 10^{-2}$ |

**Table II.** Carrier concentrations of the $Zn_{0.95}Co_{0.05}O$ thin films with different Ag concentration at 77 and 300K. The applied current was maintained at 1 μA for all measurements.

| Samples | Carrier concentration | |
|---|---|---|
| | 77 K | 300 K |
| $Zn_{0.9}Co_{0.1}O$ | $-1,6\times10^{16}$ cm$^{-3}$ | $-2.5\times10^{19}$ cm$^{-3}$ |
| $Zn_{0.895}Co_{0.1}Ag_{0.005}O$ | $-4.2\times10^{15}$ cm$^{-3}$ | $-1.2\times10^{18}$ cm$^{-3}$ |
| $Zn_{0.89}Co_{0.1}Ag_{0.01}O$ | $-5.3\times10^{11}$ cm$^{-3}$ | $-1.9\times10^{15}$ cm$^{-3}$ |
| $Zn_{0.85}Co_{0.1}Ag_{0.05}O$ | $-7.9\times10^{10}$ cm$^{-3}$ | $-3.0\times10^{14}$ cm$^{-3}$ |

**Table III.** Calculated formation energies ($E_{Form}$), magnetic moments on Co atom ($m_{Co}$), and parameter of exchange interaction ($J$) for different Co-involving defect configurations with or without presence of oxygen vacancy ($V_O$). $J_1$ and $J_2$ correspond to the exchange integral for exchanges within a defect cluster and between clusters, respectively.

| Defect configuration | $E_{Form}$ (eV) | $m_{Co}$ ($\mu_B$) | $J_1$ (meV) | $J_2$ (meV) |
|---|---|---|---|---|
| $Co_S$ | 0.03 | 4.01 | - | 0 |
| $2Co_S$ | 0.64 | 4.00 | -14 | 0 |
| $2Co_S+V_O$ | 1.04 | 3.86 | -12 | 0 |
| $Co_I$ | 2.19 | 2.97 | - | 0 |
| $Co_S+Co_I$ | 1.86 | 2.34 | 10 | 0.4 |
| $2Co_S+Co_I$ | 1.42 | 2.01 | 15 | 0.7 |
| **$3Co_S+Co_I$** | **1.07** | **1.67** | **21** | **2.1** |
| $3Co_S+Co_I+V_O$ | 1.52 | 1.72 | 20 | 1.0 |
| $4Co_S+Co_I$ | 1.23 | 1.87 | 16 | 0.8 |
| $6Co_S+Co_I$ | 1.45 | 2.15 | 13 | 0.1 |

**Table IV.** Calculated formation energies ($E_{Form}$), magnetic moments on Co atom ($m_{Co}$), and parameter of exchange interaction ($J$) for different Co and Ag-involving defect configurations with or without presence of oxygen vacancy ($V_O$). $J_1$ and $J_2$ correspond to the exchange integral for exchanges within a defect cluster and between clusters, respectively.

| Defect configuration | $E_{Form}$ (eV) | $m_{Co}$ ($\mu_B$) | $J_1$ (eV) | $J_2$ (eV) |
|---|---|---|---|---|
| $Ag_S$ | 4.30 | - | - | - |
| $Ag_I$ | 4.32 | - | - | - |
| $Co_S$+$Ag_S$ (near) | 1.82 | 3.99 | - | 0 |
| $Co_S$+$Ag_S$ (far) | 1.99 | 4.01 | - | 0 |
| $Co_I$+$Ag_S$ | 2.13 | 3.01 | - | 0 |
| $Co_S$+$Ag_I$ (near) | 2.28 | 3.85 | - | 0.1 |
| **3$Co_S$+$Co_I$+$Ag_S$ (near)** | **0.95** | **1.49** | **18** | **1.5** |
| 3$Co_S$+$Co_I$+$Ag_S$ (far) | 1.34 | 1.65 | 20 | 1.6 |
| 3$Co_S$+$Co_I$+$Ag_S$+$O_V$ (near) | 1.36 | 1.56 | 18 | 1.2 |
| 3$Co_S$+$Co_I$+$Ag_S$+$O_V$ (far) | 1.66 | 1.58 | 19 | 1.3 |
| 3$Co_S$+$Ag_I$ | 1.23 | 3.05 | 3 | 0.1 |
| 3$Co_S$+$Co_I$+$Ag_I$ | 1.84 | 1.67 | 23 | 1.1 |
| 6$Co_S$+$Ag_I$ | 1.76 | 3.05 | -20 | 0 |
| 5$Co_S$+$Co_I$+$Ag_S$ | 1.76 | 2.00 | 10 | 0 |
| 4$Co_S$+$Co_I$+2$Ag_S$ | 2.20 | 1.85 | -17 | 0 |
| 5$Co_S$+ $Ag_S$+$Ag_I$ | 2.02 | 3.06 | -14 | 0 |
| 7$Co_S$+$Co_I$+2$Ag_S$ | 1.75 | 2.14 | 6 | 0 |
| 8$Co_S$+$Ag_I$ | 1.91 | 2.14 | -10 | 0 |

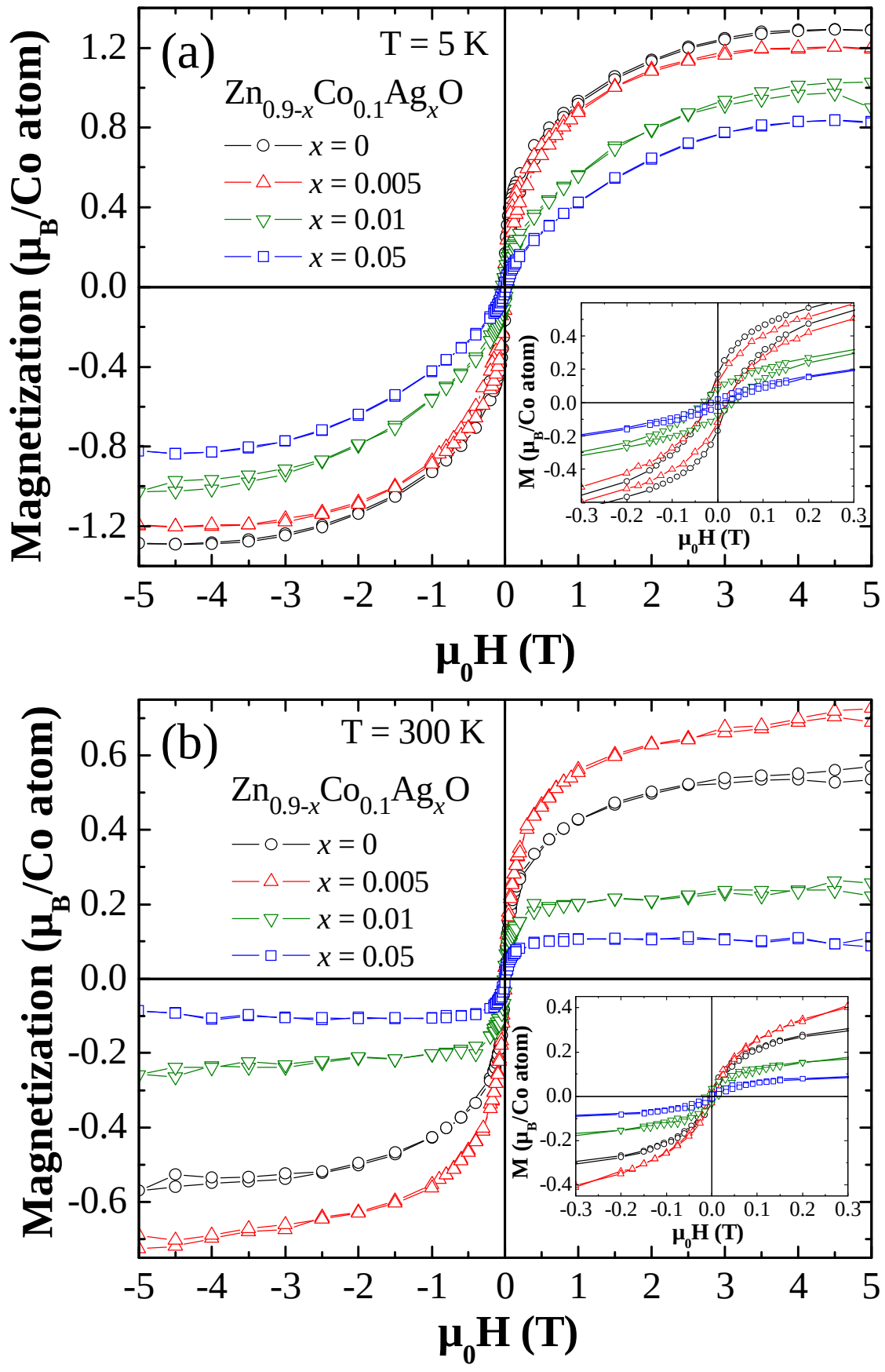

**Figure 1.** Magnetic hysteresis loops measured at 5K (a) and 300K (b) for Co:ZnO thin films doped with 0.5, 1, and 5 % of Ag. The insets show the magnified loops around zero magnetic field.

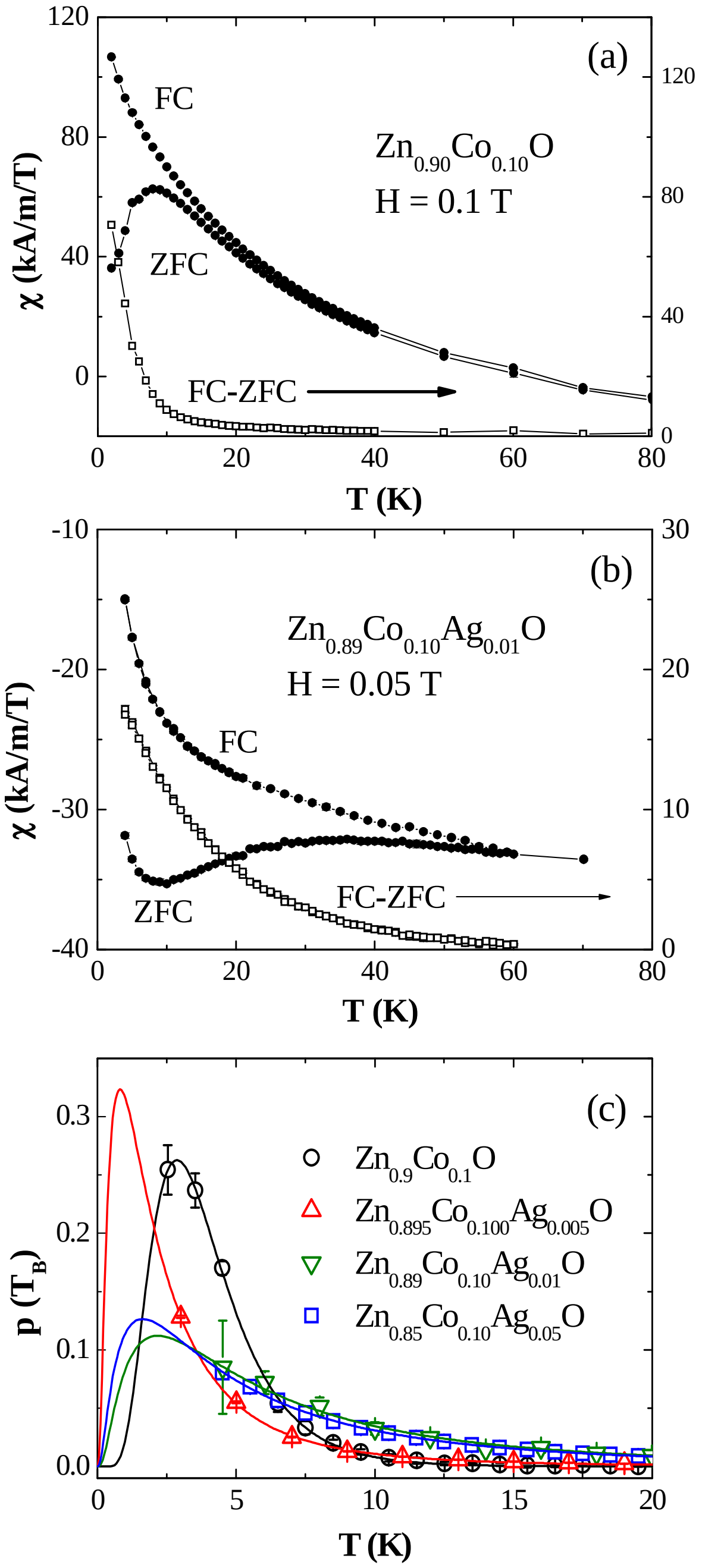

**Figure 2.** ZFC/FC magnetization curves of $Zn_{0.9-x}Co_{0.1}Ag_xO$ thin films with $x$ = 0.0 (a) and 0.01 (b), and comparison of the blocking temperature distributions in all samples (c).

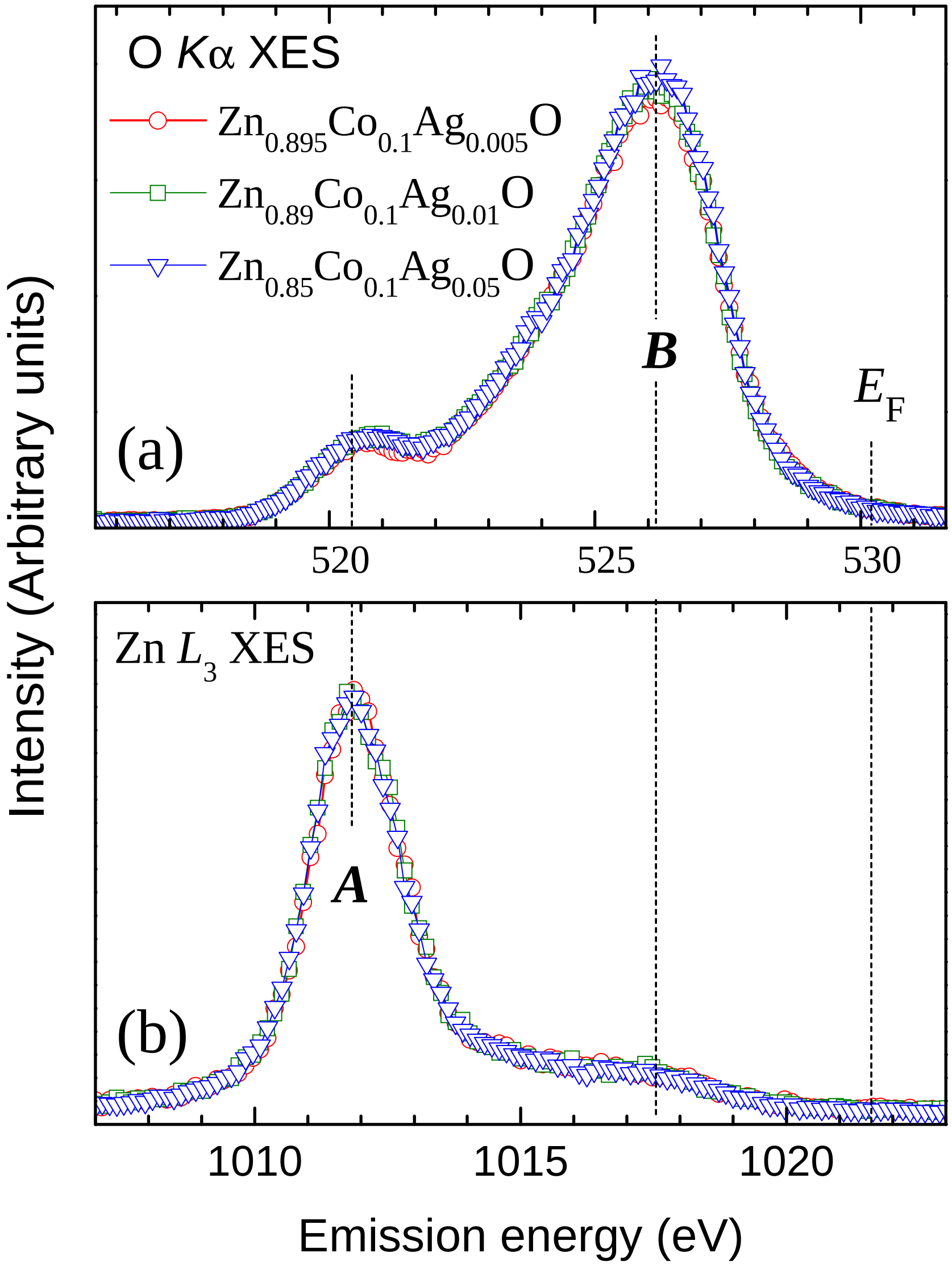

**Figure 3.** O *K*α (a) and Zn $L_{2,3}$ (b) XES spectra of $Zn_{0.9-x}Co_{0.1}Ag_xO$ thin films with different Ag concentrations.

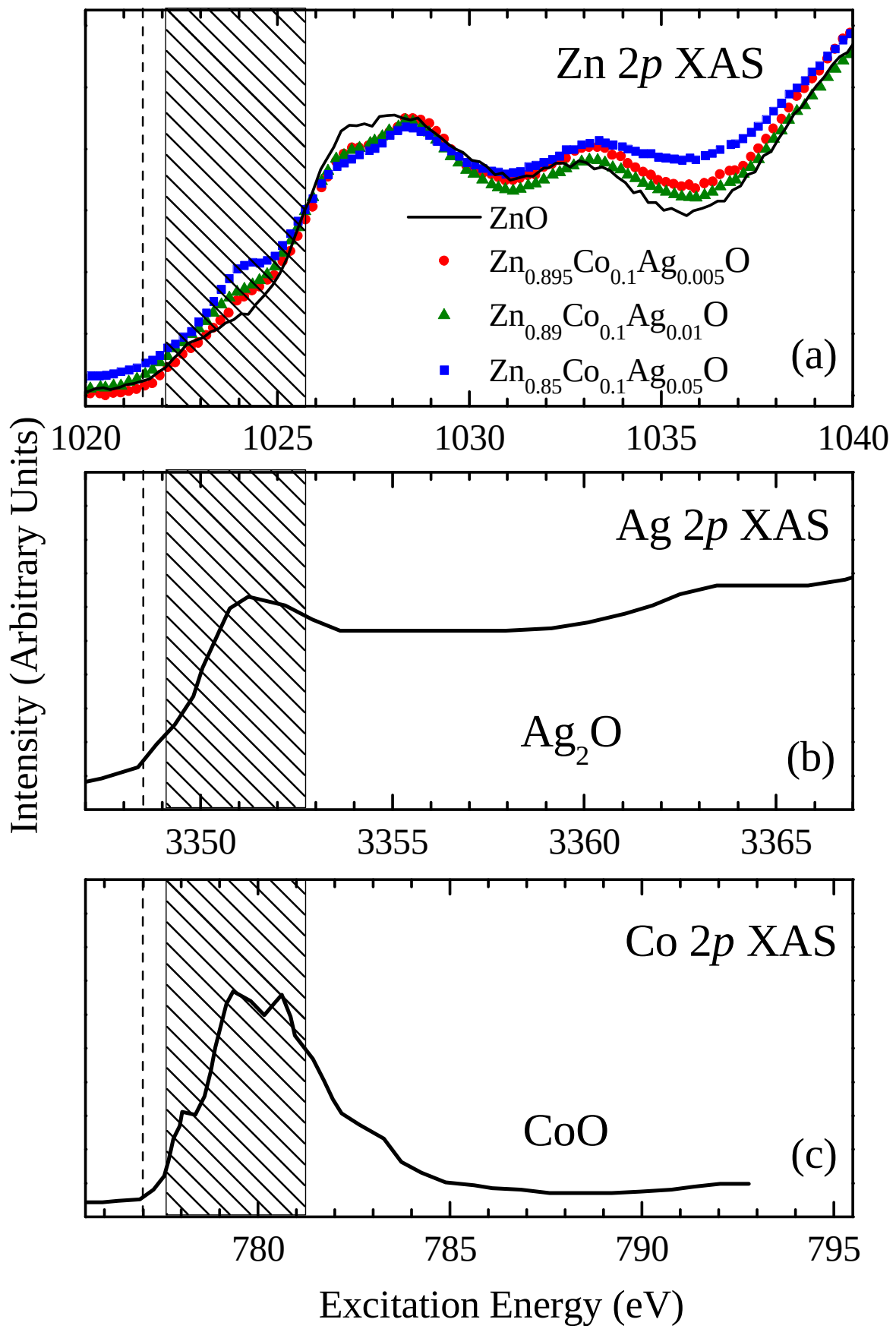

**Figure 4.** Comparison of Zn 2*p* XAS of $Zn_{0.9-x}Co_{0.1}Ag_xO$ thin films with Ag 2*p* XAS of $Ag_2O$ and Co 2*p* XAS of $Zn_{0.895}Co_{0.1}Ag_{0.005}$.

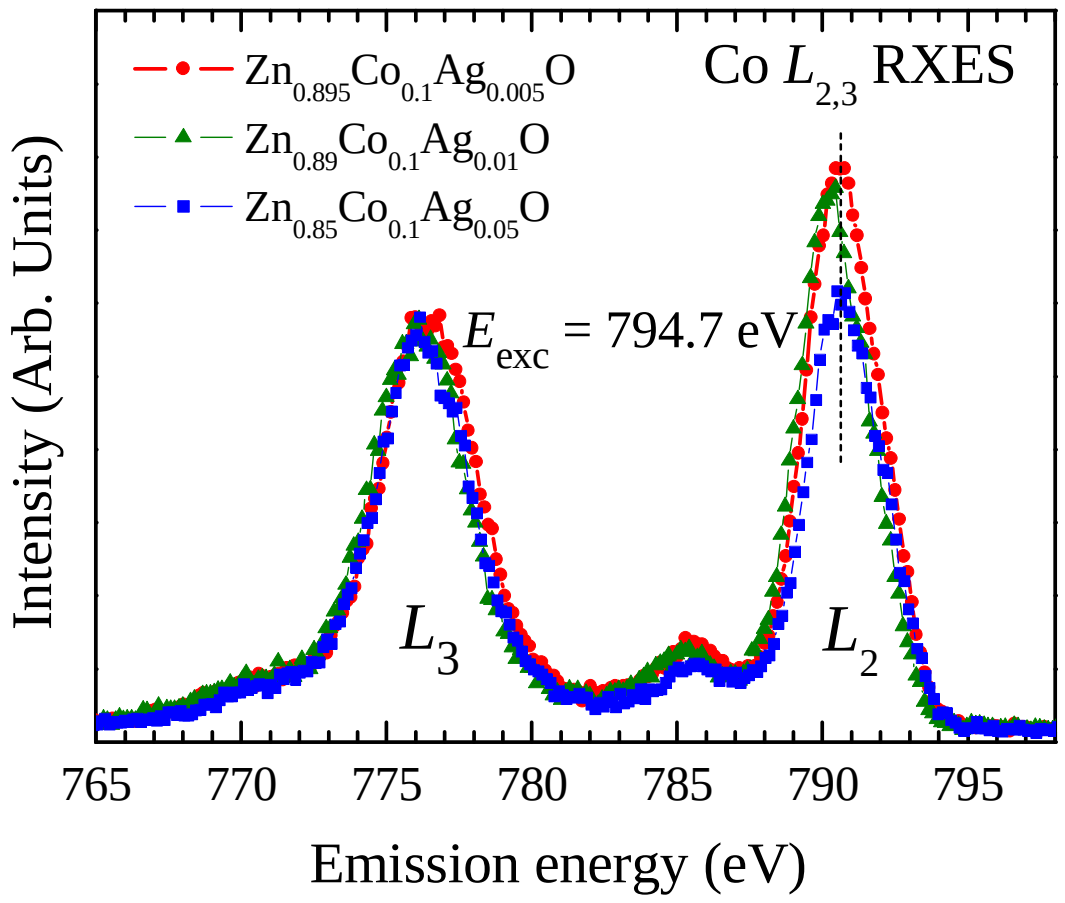

**Figure 5.** Co $L_{2,3}$ resonant XES spectra for $Zn_{0.9-x}Co_{0.1}Ag_xO$ thin films with different Ag concentrations resonantly excited at Co $L_2$-threshold.

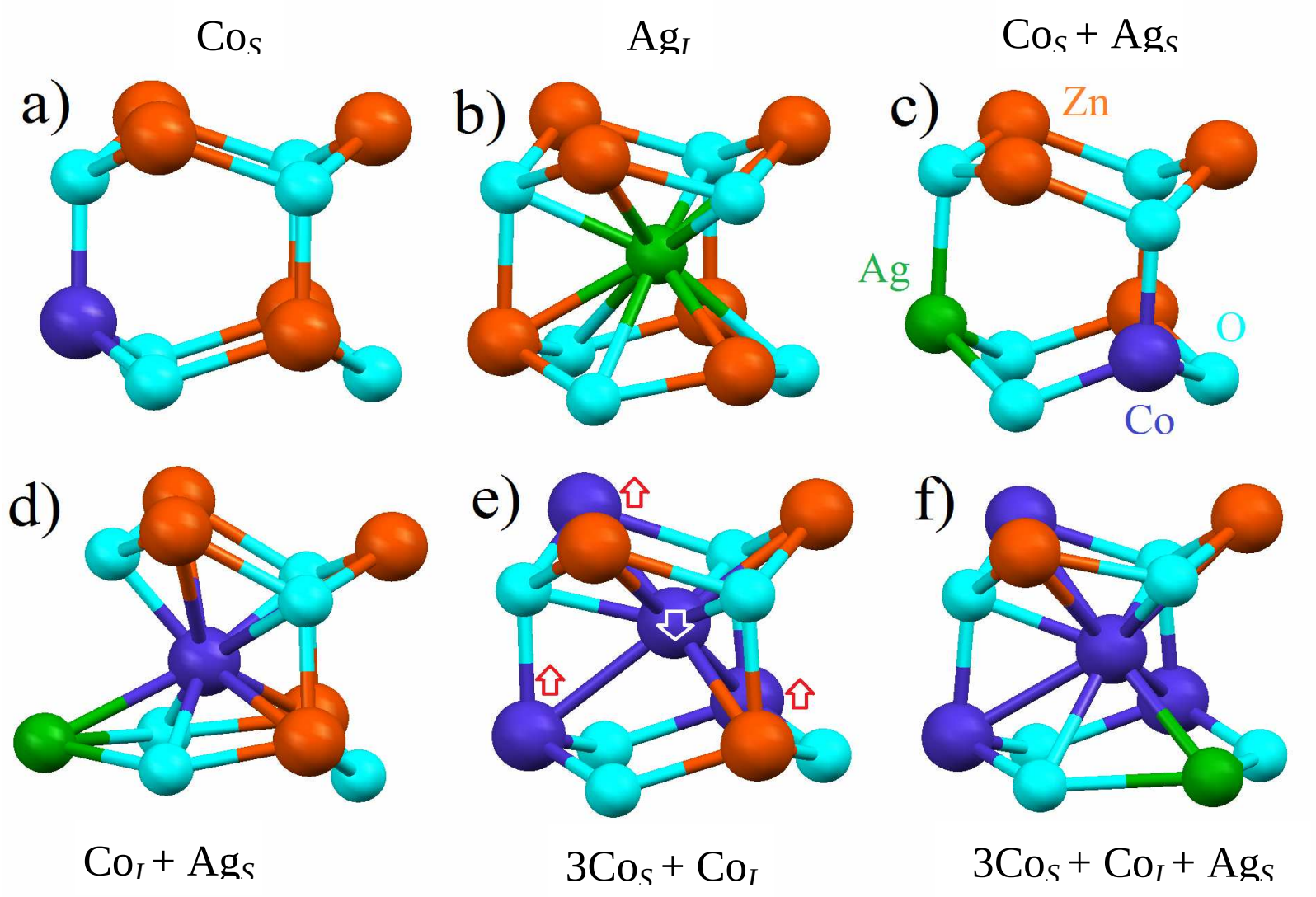

**Figure 6.** Optimized atomic structures of Zn(Co,Ag)O near the impurity atoms for configurations: (a) single substitutional Co atom for Zn site (Co$_S$); (b) single interstitial Ag atom (Ag$_I$); (c) Co and Ag atoms substituting for Zn sites (Co$_S$ + Ag$_S$); (d) interstitial Co and substitutional Ag atoms (Co$_I$ + Ag$_S$); (e) three substitutional and one interstitial Co atoms (3Co$_S$ + Co$_I$), and (f) the same structure with only one more silver atom substituting atom of zinc (3Co$_S$ + Co$_I$ + Ag$_S$). The arrows on panel (e) indicate magnetic-moment orientation of each Co atom.

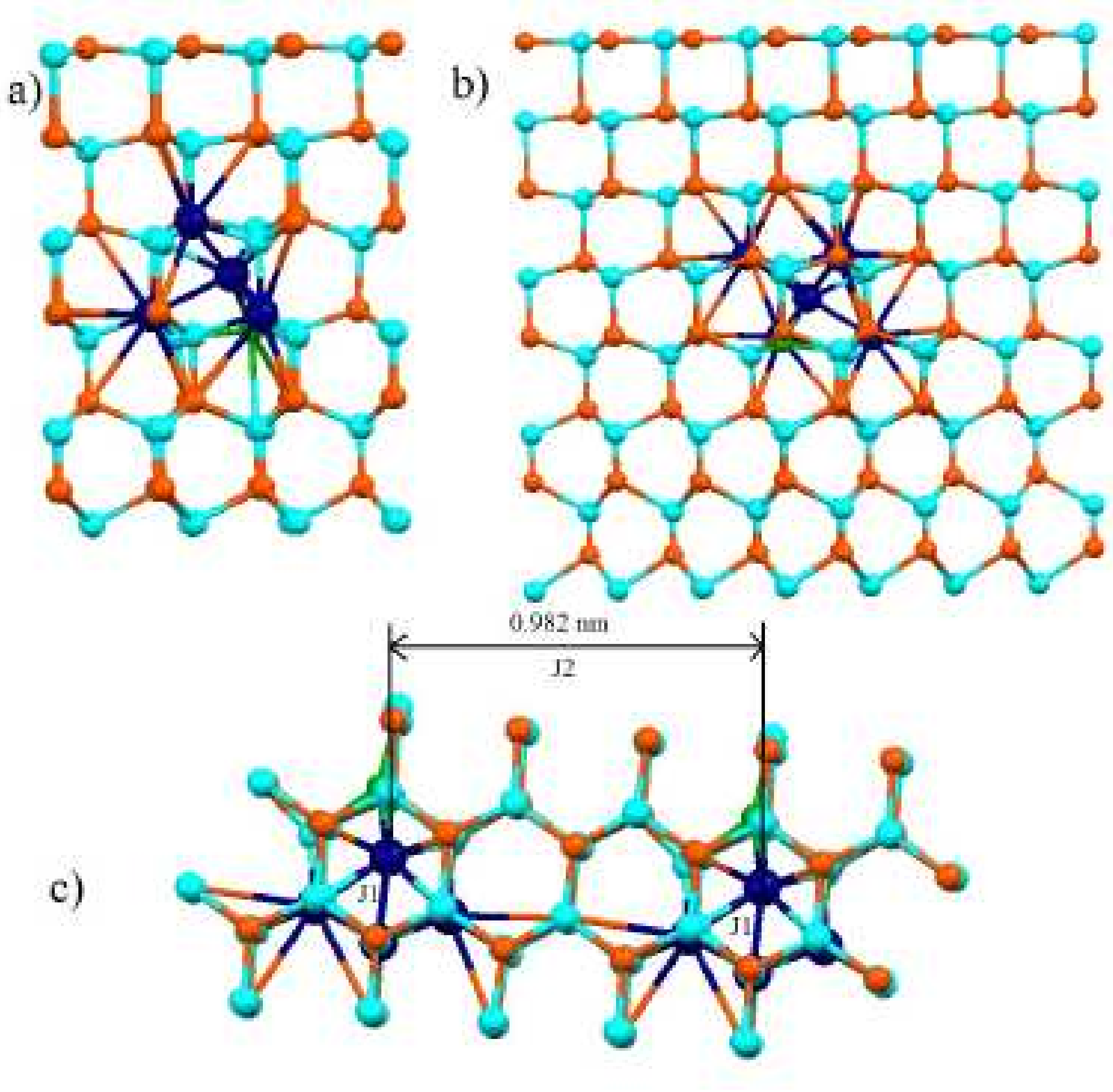

**Figure 7.** Optimized atomic structure of the supercell containing the cluster $3Co_S + Co_I + Ag_S$ with 108 atoms in the matrix of ZnO (a), optimized atomic structure supercell containing cluster $5Co_S + Co_I + Ag_S$ in a matrix of 256 atoms of ZnO (b), double supercell (along the Y axis) for the calculation of the exchange integral containing cluster $3Co_S + Co_I + Ag_S$ with 108 atoms in the matrix of ZnO (c).

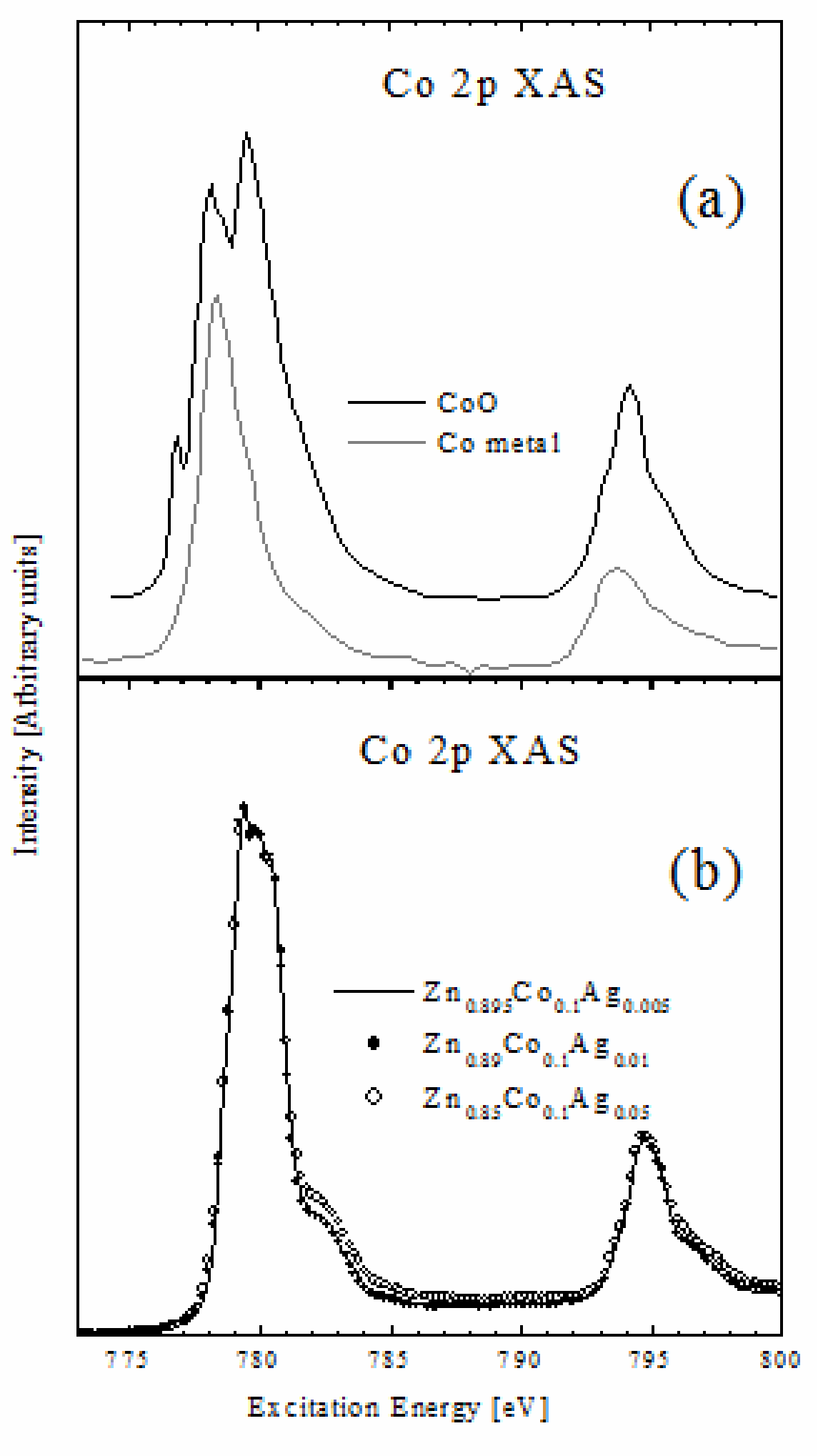

**Figure 8.** Co 2*p* XAS spectra of (a) CoO and Co metal reference samples and (b) $Zn_{0.9-x}Co_{0.1}Ag_xO$ thin films with different Ag concentrations.